# Gender identity and relative income within households: Evidence from China[*]

Han Dongcheng[†], Kong Fanbo[‡], Wang Zixun[§]

August 9, 2021

**Abstract**

How does conforming to traditional gender roles affect women's labour outcomes? To investigate this question, we use the discontinuity test and fixed effect regression with the time lag to measure how married women in China diminish their labour outcomes and maintain the breadwinning status of their husbands. In the first part of the research, the discontinuity test exhibits the missing mass of married women who just earn more than husbands, and we take it as evidence that these women are diminishing their earnings under the influence of gender norms. In the second part of the research, we use fixed effect regression with time lag to assess the change of labour outcomes in the future if a woman currently earns more than husband. Our results suggest that it will not affect women's labour participation decisions, yet women would reduce their yearly incomes and weekly working hours in the future. Heterogeneous studies are conducted to show that low-income and less educated married women are more susceptible to the influence of gender norms. JEL Codes: D10, J12, J16.

[*]We would like to thank Ms. Li Hongyan of the National University of Singapore and Mr. Adams Hu of the Cornell University for their valuable comments and guidance.
[†]Han Dongcheng (Dunman High School, Singapore)    Email: han.dongcheng@dhs.sg
[‡]Kong Fanbo (Nanyang Junior College, Singapore) Email: kong.fanbo_2021@nyjc.edu.sg
[§]Wang Zixun (River Valley High School, Singapore) Email: wang_zixun@rvhs.edu.sg

# Contents

# 1   Introduction

The female labour force participation (FLFP) rate has been persistently dropping for the last three decades in China (ILO, 2021). If such a trend continues, less than 3 out of 5 women will participate in the labour force in a few years. What leads to the shortage of female labour supply? In addition, according to Urban Household Survey Data by National Bureau of Statistics, the average annual earnings of females is persistently lower than males by 20%. One might ascribe this phenomenon to the "glass ceiling" in China's labour market, suggesting that there are invisible gender biases existing in workplaces in payment, promotion opportunities and welfare, discouraging females from workplaces. However, such an explanation might be incomplete, as the declining trend of FLFP rate persists even though the central government emphasised the importance of protecting working females' legal rights and enacted the law to prohibit discrimination against women in the labour market. We suspect that the declining trend of FLFP rate and persistent wage gap are caused by females distorting their labour outcomes under certain social norms.

Identifying what leads women to distort their labour outcomes is critical because an imbalance in the supply of male and female labour force could have severe consequences. Fewer women participating in the labour market tends to decrease the overall productivity and will impede economic growth (Bertay et al., 2020; Bandiera & Natraj, 2013; Klasen, 2018). The severe consequences are aggravated in China because China is facing a labour force shortage in recent years. Furthermore, if a woman decides not to work, it might affect the labour decision of her subsequent generations. Fernández et al. (2004) confirm that a woman's employment status positively and significantly correlates with whether her mother-in-law worked in the past. Furthermore, they propose a mechanism that a man whom a working mother raises is more likely to support his wife to participate in the labour market. This relationship is corroborated by X. Chen & Ge (2018). Using the data in China, they point out that a declining female participation rate means more boys have non-working mothers, and when becoming adults, they are less likely to support their wives' work. This precipitates the shortage of female labour supply in the long run. Therefore, recognising the causes for women's reduction in labour supply has strong policy relevance in China.

A multitude of explanations have been given by researchers. Maurer-Fazio et al. (2010) and Shen et al. (2016) explain that the presence of younger children forces females to reduce their labour outcomes as women are traditionally assigned the responsibility of looking after their children. Such an effect is intensified for those adults not residing with their parents (who could share the burden of childcare). Y. Chen et al. (2020) employ personnel records of a large company to illustrate that women start to work less and decline promotion opportunities after marriage and becoming parents, resulting in a widening gender pay disparity. Li et al. (2006) argue that the economic transition from a planning economy to a market economy gives households more liberty to allocate production duties among household members as the government no longer directly assigns jobs to people. Division of labour between spouses occurs such that husbands become the main financial supporters while wives work less in the labour market and devote more time to housework. Such a phenomenon is also confirmed by de Bruin & Liu (2020). Furthermore, it is noted in their paper that women from households in rural areas of China spend more time on unpaid housework and less on paid work in the market than the women in urban ones. Ye & Zhao (2018) document that women produce less economic outcomes in the labour market if their husbands agree that family is women's primary responsibility. Seemingly, all these explanations point to one question: what roles and responsibilities does a woman have?

According to the World Health Organisation (WHO, 2015), women's roles and responsibilities in a society could be referred to as gender roles. In China, the society generally embodies more traditional and less egalitarian views to gender roles, and surveys and so- ciology research manifest this. Approximately 40% of Chinese respondents in the World Value Survey (WVS) agree that "Wives should not earn more than husbands" and 60% of respondents agreed that "Being a housewife is a fulfilling thing" (WVS, 2017). Further- more, 2 out of 3 respondents in the China General Social Survey (CGSS) agree that "a male should prioritise career while a female should prioritise family" and approximately 40% of respondents agree with the statement "during the recession, female workers should be dismissed first" (CGSS, 2013). These survey results confirm that the Chinese society mainly recognises men as the major bread-maker within a household. Sociology articles have also substantiated the existence of traditional gender role beliefs in China. McKeen (2005) argues that Chinese male students expect their wives to do more domestic work and have less prestigious jobs and less earnings than themselves. Pimentel (2006) points out the trend that more Chinese men seem to embody less equal views to gender roles.

There are multiple attempts to explain the origin of the less equal gender views in China, three of which are relatively credible: Alesina et al. (2013) point out that de- scendants of societies that previously practised plough agriculture could have less equal gender views. In ancient times, physically demanding plough agriculture was mainly conducted by males. This created a division of labour along the gender line where hus- bands specialised in working in the field while wives produced at home. As a result of the two-thousand-year practising plough agriculture, a gender norm that husbands are mainly responsible for bread-winning has been shaped and persistently influences the labour market in modern society in China. Another explanation relates to the main- stream ideology in China. The mainstream ideology stems from Confucianism, which advocates that wives should be unconditionally obedient to their husbands' instructions. Any attempt from wives to oppose their husbands' decision would be deemed as inap-propriate and deserving moral condemnation. Husbands, therefore, attained dominant roles in household economic decisions (Rosenlee, 2012). Last but not least, Shu (2004) emphasises the significance of education in advocacy of gender egalitarianism, arguing that less educated people could hold less equal gender views. This explanation is testified by the fact that in WVS, less educated respondents are more likely to agree with less gender-equal statements. In China, the average schooling year is 8.7 (UNDP, 2021), and this number could be much lower in rural areas. Less access to education in rural areas therefore should be a reason for the less equal gender views.

The gender role beliefs in China could be summarised as follows: males are the main breadwinners, and females are the main homemakers. Wives are not encouraged to gener- ate more productive economic outcomes than their husbands because it is not their main job. Akerlof & Kranton (2000) argues that individuals have an incentive to adhere to a social custom when the loss of reputation could otherwise sanction them. This forces women to conform to the traditional gender role beliefs and distort their labour outcomes to ensure they earn less than their husbands.

To quantify the how traditional gender views affect the labour market, Bertrand et al. (2015) establish an innovative analytical framework. They confirm the existence of such a gender norm that "wives should not earn more than husbands", which influences women's relative income, home production, marriage satisfaction and divorce decision. They first employ the McCrary (2008) test on the point where the wife's relative income is ½ and exhibit a sharp drop of distribution on the right of that point. To explain this discontinuity, they write that wives who are able to earn more than husbands are influenced by the gender norms and reduce their incomes. They go a step further to identify that women with higher chances of earning more than their husbands are more likely to earn below their potential, decrease working time, and increase their time on housework. Their work kick-started discussions on how gender norms could impact labour outcomes in different countries. The conclusion of Bertrand et al. (2015) is verified in West Germany (Wieber & Holst, 2015) where the society embraces a less gender-egalitarian view. Sprengholz et al. (2020) further confirm that married women in West (but not East) Germany diminish their labour outcomes to avoid the situations where they earn more than their husbands. They also find that the significance of male breadwinner prescription decline in West Germany since the reunification, converging to more gender-egalitarian East Germany. Their work suggests that the significance of male breadwinner prescriptions is not static but rather subject to the change of political and institutional environment.

In our research, we employ the data from China Household Financial Survey (CHFS, 2020). Our empirical model is mainly based on Bertrand et al. (2015). We begin our re- search by conducting the McCrary (2008) test to demonstrate that there is a discontinuity in the distribution of wife's income share to the right of the ½ point. Our results suggest that the magnitude of discontinuity is larger than those obtained in Western countries (Bertrand et al., 2015; Wieber & Holst, 2015). According to Bertrand et al. (2015), the discontinuity demonstrates that working married women are influenced by gender norms and they diminish their income to avoid earning more than their husbands. We divide our samples into different groups and conduct McCrary (2008) tests to explore the het- erogeneity. Also, we notice that a strand of literature disagrees with this identification strategy(Binder & Lam, 2020; Hederos Eriksson & Stenberg, 2015; Zinovyeva & Tver- dostup, 2018). They argue that the discontinuity is driven by the mass point of spouses with identical income, and they provide an alternative explanation that husbands and wives equalise their income by working together with each other. To address these con- cerns, we augment our identification strategy to make the gender norm explanations more robust. We conclude that spouses' income equalisation is a less plausible explanation for the spike at ½ in the distribution of wives' relative income. We therefore ascertain that gender norm is the main driving force to women's distortion of labour outcomes, and we use the second section to substantiate this idea.

In the second section of our research, we build a model that predicts wives' response in their future labour outcomes when their current incomes exceed their husbands'. We hypothesise that wives who earn more than husbands could diminish their labour outcomes in the following ways: leaving their employment, reducing their wages or reducing their working hours to conform to gender norms. We employ fixed effect regression with time lag to test our hypothesis. A basket of demographic characteristics are controlled, and results are statistically significant. To inquire into heterogeneity, we explore how the impact of traditional gender views vary across groups with different educational attainment and socio-economic status.

This research mainly contributes to two strands of literature. The first strand of lit- erature debates the interpretation of mass point at ½ in the distribution of wives' relative income. Bertrand et al. (2015) start the strand by interpreting the mass point as the con- crete evidence of women reducing their income to restore gender prescription. The same identification strategy is used in Germany and Canada (Wieber & Holst, 2015; Sprengholz et al., 2020; Doumbia & Goussé, 2019). However, such identification strategy has not been used in China before our investigation. On the flip side, their critics (Binder & Lam, 2020; Hederos Eriksson & Stenberg, 2015; Zinovyeva & Tverdostup, 2018) argue that this identification strategy is problematic because the discontinuity could be consequences of equalisation of spousal income after marriage. Binder & Lam (2020) suggest that self- employers and spouses working together tend to earn identical income, which causes the discontinuity, which is further substantiated by Zinovyeva & Tverdostup (2018). Their work shows that the discontinuity could be largely explained by the convergence of income due to husbands and wives working on the same premise, and the convergence occurs both in the group of spouses where husbands earn more than wives and in the group of spouses where the wife earns more than her husband. On the basis of that, they reject the con- clusions of Bertrand et al. (2015) in Denmark. We contribute to this strand of literature by validating the conclusions of Bertrand et al. (2015) in China's context. Our results suggest that the discontinuity at the right of ½ in the distribution of wives' relative income is still mainly driven by gender norms. Convergence of income due to spouses working together contributes to the spike but is less likely to be the main driving force. We note that co-working spouses also succumb to gender norms in China. Furthermore, we also explore the how the magnitude of discontinuities vary across different income cohorts.

Our investigation also relates to the strand of literature that evaluates social norm's impact on the supply of labour both in China and worldwide. Different types of social norms that are less gender-egalitarian - including culture, religion, gender role attitude are proven to impact the labour supply of women. Johnston et al. (2014) indicate that a mother's non-traditional gender role attitude could increase the supply of labour and human capital on their daughters and daughters-in-law, especially after their childbirth. Scoppa & Stranges (2019) opine that a female immigrant's decision on labour participation in Italy is mainly dependent on the female labour participation rate of her mother country. Cavapozzi et al. (2021) argues that having peers with more gender-egalitarian views could increase a female's probability of having a paid job. In China's context, the impact of husbands' and wives' gender identity on wife's labour supply has been investigated by Ye & Zhao (2018). Their work shows that the impact of gender identity varies across generations, regions and education levels. Most importantly, the husband still "guides" his wife's labour outcome because his gender identity exerts a stronger impact than her gender identity. X. Chen & Ge (2018) study how gender identity could transmit across household members and affect labour outcomes. They present evidence of a positive correlation between the labour force participation probability of a married woman and the former working experience of her mother-in-law. This strand of literature is limited in three ways:
(1) Some of them (X. Chen & Ge, 2018; Scoppa & Stranges, 2019) embody a narrow definition of labour supply that they only target on how women's labour participation decision is affected by gender norms. In reality, it is possible that married women still work yet tend to reduce their income with the impact of gender norms. (2) This strand of literature mainly studies the static relations between women's labour outcomes and gender identity but does not fully reveal spouses' dynamic response to infringement to gender norms. (3) This strand of literature does not adequately address the heterogeneous impact of gender identity. Our investigation could be added to this strand of literature because we employ a broader definition of labour outcomes. Our dependent variables include women's labour force participation decisions, the change in yearly income and the change in weekly working hours. Our model suggests a dynamic adjustment process where wives who currently out-earn husbands will adjust their income and earn less in the future to conform to traditional gender views. We further find out the heterogeneity that low-income and less educated married females are the most susceptible to the impact of gender identity.

The rest of the paper is organised as follows. Section 2 provides an overview on the theoretical framework and the labour market in China. Section 3 gives a description on the dataset. Section 4 articulates the results and interpretations of the McCrary test. Section 5 presents the regression analysis. Section 6 is the robustness check. Section 7 concludes.

# 2  Background and Theoretical Framework

## 2.1  Identity economics

Our theoretical model is based on Akerlof & Kranton (2000) which considers the effect of identity on economic outcomes. On the basis of that, we argue that the gender role that describes what is appropriate and inappropriate for females and males could cause males and females to alter their labour outcomes. Akerlof & Kranton (2000) argues that the identity of individual $j$ depends on the social categories he falls into. The social categories could be gender and race, which are fixed and predetermined. In each social category, a set of prescriptions defines what is appropriate for members of that social category to behave. In their designed setting, Akerlof & Kranton (2000) argue that utility of individual $j$ depends on her own action $a_j$, other people's action $a'_j$ and the identity $i_j$, where

$$U_j = f(a_j, a'_j, i_j) \tag{1}$$

In addition to the vector of action $a_j$ and $a'_j$, $j$'s identity $i_j$ also could be influenced by the social categories $c$ and prescription $P$ that defines how members in that category should behave, as well as the extent of matching to the characteristics of social category $e$. In sum, this could be formalised as:

$$U_j = f(a_j, a'_j, i_j, c_j, e_j, P) \tag{2}$$

In the context of our research, we assume binary gender identities where each married individual defines himself or herself clearly as male or female, and he or she would max- imise the utility by conforming to the prescription of the assigned social category. We argue that the gender prescriptions are as following: (i) *Males should earn more than females*. This is evident by World Value Survey: approximately 35% of respondents in China agree with the statement that "there will be problems if women earns more than female" (WVS, 2017). Akerlof & Kranton (2000) argue that it is a common phenomenon that husbands expects themselves to earn more than their wives, and they might feel uncomfortable if their wives out-earn them. (ii) *Males are the main breadwinners while* *females are the main homemakers.* This is supported by evidence from the World Value Survey where around 68% of respondents agree that "being a housewife is as fulfilling as working outside" In addition, numerous researchers have proved that many women re- duce their working hours and decline promotion opportunities out of concerns that their primary responsibility is to take care of their families (Lundborg et al., 2017; Y. Chen et al., 2020; Li et al., 2006).

Based on the prescriptions above, we argue that females would conform to the gender prescriptions to maximise their utility. Thus, they might try to control their income and prevent it from exceeding their husbands'. Wives will attempt to restore the equality of utility function that they spend more time on housework and reduce their working hours to assuage their husbands' unease (Bertrand et al., 2015).

## 2.2 China's labour market

China's serial economic liberalisation has kick-started her robust economic growth and improvement on material living standards since 1978. But interestingly, it accompanies a widening gender gap both in terms of women's labour participation and wages earned. As estimated by the International Labour Organisation (ILO, 2021), the female labour force participation ratio decreased from 80% in 1990 to 67% in 2019. Similarly, according to World Bank data, while female workers constitute an increasing portion of the total labour force in the United States, Chinese women seemingly start to take a step back in the labour market as their weight-age in total labour decreases over the years(WorldBank, 2020). A study from Peterson Institute for International Economics (Zhang & Huang, 2021) suggests that there is a causal relationship between China's transition from a planned economy to a market economy and the widening gender gap in the labour market.

On the demand side, a market economy no longer assigns jobs and allows private and state-owned enterprises to recruit for themselves. Enterprises take advantage of their autonomy in recruitment in a competitive economic environment and discriminate against females because they believe women's productivity could be affected by their roles as homemakers. This could be substantiated by a recent study (Y. Chen et al., 2020) that shows female employees will divert their focus from working to family, causing working hours and willingness to promote. Another study led by All-China Women's Federation in 2015 found that more than 80% of female university graduates in China have experienced some form of gender discrimination in their job search, regardless of the nature of jobs (WSIC, 2016).

On the supply side, households internalise economic transformation's shock to social fabrics and re-assign the role of breadwinner and homemaker within a household. Li et al. (2006) argue that upon the economic transformation, households are free to allocate time, resources and human capital investment to each family member efficiently. This restoration of production could be attributed to two factors that the economic transition gives rise to. First, the nation no longer assigns jobs directly to its citizens and whether to be employed becomes a choice of individuals or households. Second, the government does not offer childcare services, which previously aimed to spare parents from looking after their children in their working time. Therefore, households need to plan how to assign the roles of breadwinners and homemakers among its members, and commonly women choose to take a step back in their careers. Ji et al. (2017) thus pointed out that when striking a balance between breadwinner and caregiver, women are more likely to reduce their time spent on work. Guvenen et al. (2020) also documented the gender pattern of time usage in the post-economic-reform era that women spend more time on non-paid jobs than men. As a result, women might distort their labour outcomes in three ways: they leave employment, reduce their working time, or choose to earn less. (Y. Chen et al., 2020)

A considerable body of literature emphasizes that gender inequality and imbalance in the labour market could hinder economic growth due to misallocation of resources and decreased productivity.(Bertay et al., 2020; Alesina et al., 2013; Klasen, 2018) In China, gender inequality's impediments to economic growth is more salient because China also experiences an ageing population and shortage of labour force nationwide. According to China's Department of Human Resources and Social Security, the job-to-seeker ratio has been higher than one since 2013 (MOHRSS, 2013). In addition, the total labour force declined from 787 million in 2017 to 785 million in 2018 for the first time (ILO, 2021). These could contribute to the recent slowdown of economic growth in China, and the government is cognisant of the significance of encouraging more women to participate in the labour market and supplement the shortage of labour. Hence, the government has recently made efforts in elevating women's status and preserving women's rights in the labour market: on the legislative ground, legislators specifically enacted a chapter of "equality of employment" in the Employment Promotion Law of the People's Republic of China to advocate for gender equality in the labour market in 2007. Furthermore, in 2012, the State Council implemented Special Rules on Protection of Female Employers, which legally extend female's maternal leave from 90 days to 98 days and effectively make the workplace more friendly to women. Also, in its white paper "Equality, Development, Share: Advancement and Progress of Women's Career in New China," the government acknowledges that women are playing a salient role in holding up half of the sky in economic construction (PRCGOV, 2020). However, despite these efforts, the declining trend of female's labour participation rate is not reversed, not to mention the decrease of total labour force. This necessitates further research to better understand the impediments on women's labour participation.

# 3  Data

## 3.1  Data description

In this research, our main source of data is the China Household Financial Survey (CHFS). CHFS is conducted by the Survey and Research Center for China Household Finance in the Southwestern University of Finance and Economics. The survey collect nationally representative data by interviewing the households across China biannually. The objec- tive of CHFS is to collect micro-level financial information about Chinese households for researchers to understand China better (CHFS, 2020).

Since 2010, five waves of surveys have been conducted. In this paper, we employ the survey results between 2011 and 2017. Until we finish this work, the latest survey result in 2019 has not been released yet. We note that there is a one-year delay between completion and publishing for each wave of the survey. For example, the survey result released in 2017 was completed in 2016. In our data analysis, we have added two restrictions on the CHFS original dataset. First, only couples who are both aged between 18 and 65 are kept in our dataset. Second, we restrict our sample to dual-earning couples who have both reported their specific income. Finally, we are left with 21069 observations[*] on approximately 14000 households, and this dataset is then used for the McCrary test. We further remove all households that do not participate in two consecutive surveys, and around 12000 observations are left when proceeding to the regression part.

Our dependent variable, $FemaleLabourOutcome_{i,t}$, is a placeholder for the employ- ment status, yearly income after tax, and weekly working hours of female $i$ at time $t$. The employment status of female $i$ is represented by a binary variable "$Employed$" which equals to 1 if she is employed for the past month or 0 otherwise, while her yearly income after tax and weekly working hours are numerical values reported by this respondent. Note that households with missing values for income and working hours are excluded from our dataset. Furthermore, choosing the price level in 2016 as a benchmark, we ad- just the income according to the inflation rates between years, which are collected from the National Bureau of Statistics of China.

To determine female $i$'s income as a percentage of the total income of her family, we generate a variable named $Wife\_Relative\ Inc_{i,t}$, which is calculated by dividing wife's income by the sum of spousal income. Based on $Wife\ Relative\ Inc_{i,t}$, we further define three additional variables to directly represent whether female $i$ earns more than her hus- band. When $Wife\ Relative\ Inc_{i,t}$ exceeds 0.5 (female $i$'s income is larger than 50% of her total family income), $WifeEarnsMore_{i,t}$ is coded as 1, and 0 otherwise. This variable is the main variable of interest in our research. Similarly, we can obtain $WifeEarnsEqual_{i,t}$ and $WifeEarnsLess_{i,t}$.

---

[*]An observation is an individual-year pair.

Five variables are included as the demographic control, which are: age, the square of age, affiliation to the Chinese Communist Party, education level and region of residency (hukou).

## 3.2 Data summary

We summarise the data in the following tables. Table A1 presents the variables of interest for McCrary test among all couples, while Table A2 presents the variables of interest and demographic controls in regression analysis.

From Table A2, it can be observed that around 17% of married women earn more than their husbands and approximately 4% of married women who previously reported income to become unemployed. On average, married women work 42 hours per week and their yearly income after tax is around 30,000 RMB, adjusted upon the price level in 2016.

# 4 McCrary Test

McCrary (2008) develops a methodology to test the continuity assumptions on running variables in regression discontinuities. The proposed test is based on an estimator for the discontinuity at the cutoff in the running variable's density function. The estimation consists of two steps. First, a gridded histogram about the running variable is plotted. Second, two local non-linear functions are used to smooth the density function separately on both sides of the cutoff. The discontinuity is specified as:

$$\hat{\theta} = ln(\hat{f}_+) - ln(\hat{f}_-) \tag{3}$$

We choose to employ this methodology because the discontinuity in the density function of wife's relative income is of our primary interest. Note that wife's relative income is calculated by dividing wife's yearly income after tax by the total spousal income after tax in the given year.

Bertrand et al. (2015) argue that gender role beliefs pressure married women into diminishing their earnings to avoid out-earning their husbands, resulting in a regression discontinuity at the 0.501 point of wife's relative income — a point where a wife starts to earn more than her husband. Furthermore, they assert that such a phenomenon is preva- lent, leading to the missing mass of married women who just earn more than husbands. A discontinuity hence arise at the right of the 0.5 point (referred as 0.501 point hereafter). Bertrand et al. (2015) use data from 1990-2004 Survey of Income and Program Participa- tion (SIPP) and estimate that discontinuity at 0.501 point is 12.3% ($p < 0.1$). Sprengholz et al. (2020) document that such a discontinuity decreases in West Germany from 55% between 1984 and 1990 to 11% between 2007 and 2016. Other than that, Doumbia & Goussé (2019) exhibit a smaller discontinuity of approximately 5% using data of 2006 and 2016 Census data in Canada's context.

We apply the McCrary test on CHFS data. By restricting our samples to dual-earning spouses who report their yearly income after tax in specific numbers, we have 13115 observations in total. We set the female relative income as the running variable, and breakpoint at 0.501. Our baseline result suggests that the estimated log difference on both sides of the 0.501 point is -2.28 ($p < 0.01$). This value is larger than any value obtained in the United States, West and East Germany, and Canada. This could mean that many wives in China who earn slightly more than their husbands diminish their earnings to achieve equal earnings between spouses. Thus, we can deduce that the gender norms in China incentivise wives to diminish their labour outcomes.

Table 1: **Density Discontinuity Estimates for Working Couples (Including and Excluding Equal Earning Couples)**

| Group | Log Diff. | Obs | SE |
| --- | --- | --- | --- |
| All dual–earning couples | -2.28 | 21,069 | 0.05 |
| Non-identical earning couples | -0.33 | 13,072 | 0.06 |

Note: We employ McCrary test to find whether there is a discontinuity at the right side of the spousal equal earning point (referred as 0.501 point hereafter). The discontinuity suggests a missing mass of couples where wives just earn more than husbands, which indicates wives could have distorted their income to avoid their incomes exceeding their husbands'. We carry out the test on the data from China Household Financial Survey (CHFS). The discontinuity is large and statistically significant. We also omit all the equal earning couples because their coincidence of income could be a result of manipulation. After removing these couples, the discontinuity still exists. Figure 1 and Figure 2 demonstrate our results.

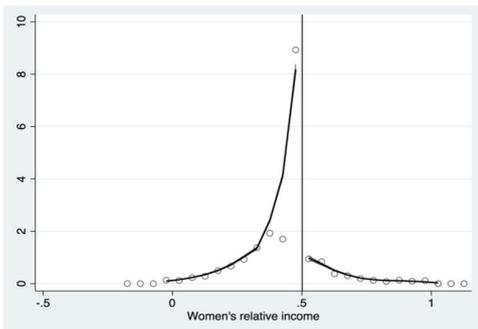

Figure 1: McCrary test for all couples at the right side of the spousal equal earning point (0.501).

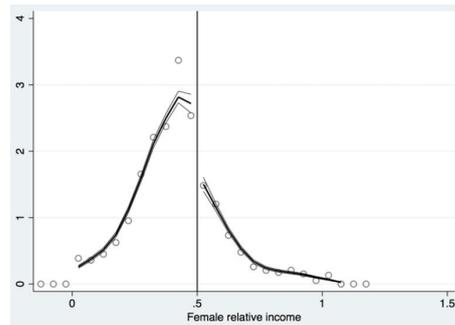

Figure 2: McCrary test for non-identical earning couples at the right side of the spousal equal earning point (0.501).

Binder & Lam (2020) disagree with this identification strategy. They suggest that this discontinuity could also be explained such that wives who earn slightly less than their husbands attempt to earn more to achieve equal income. They test their proposition by omitting the 0.5 spike, and the discontinuity drops from 12.3% to 3.3%. A similar robust- ness check is also used in the analysis of Sweden (Hederos Eriksson & Stenberg, 2015).

Furthermore, they suggest that most of equal earning couples are both self-employed individuals (including couples running business together), so they simply divide their yearly household income by two and submit it as their personal ones. In our dataset, only the employees are asked to report their yearly income after tax. Even if we exclude all the couples with identical earnings, the regression still exhibits a clear and significant dis- continuity (33%) at the 0.501 point. The existence of discontinuity exhibits that there is a missing mass between females who just earn less than husbands and females who just earn more than husbands. This is supportive evidence that married females purposefully reduce their earning and avoid the scenario that they earn more than husbands.

Hence, we conclude: the discontinuity in 0.501 point could evidently speak for how gender identity distort married women's labour outcomes. The spike at the 0.5 point of wife relative income arises not because of coincidence of spousal income, but because of wives manipulating their incomes and avoid threatening the breadwinner status of husbands. By comparison, other explanations are inadequate in addressing why the dis- continuity still exists after omitting all equal-earning couples. Hence, our explanation of gender norms could be the most suitable one for the discontinuity at the 0.501 point.

The magnitude of discontinuity could be a proxy of the impacts of traditional gender role beliefs, as a larger discontinuity indicates more married women diminish their income to conform to gender norms. Therefore, we can investigate how traditional gender role beliefs impact various groups differently by looking into the differences in the magnitudes of discontinuity.

We group the couples based on the wives' earnings into ten deciles. The variation of the number of observations in different deciles is due to the coincidence of mass points and faulting lines. In each decile, only the mass point on the upper limit is included. Thus, approximately each income group consists of 2,100 couples. We then employ the McCrary test at the right of identical earning points (0.5001) on all groups and A3 shows the results. It can be observed that the magnitude of discontinuity decreases as wives' earnings ascend across income groups. We visualise our results in Figure 3 and 4. Women with higher income are arguably less susceptible to the impact of traditional gender role belief, and women with lower income reduce their earnings due to gender norms. The correlation between women's earnings and gender norms' impact is essentially linear.

However, when we group the females into ten deciles based on their household income, the correlation between income percentile and discontinuity becomes non-linear according to Table A4. The visualisation results in Figure 5 and 6 suggest that an inverted U-shape could fit the pattern between income percentile and discontinuity.

We offer two possible explanations to address the heterogeneity. Firstly, education might explain why high-income females are less susceptible to gender norms and less likely to diminish their income. A body of literature has testified the role of education in promoting gender egalitarian views. Boehnke (2011) recounts that individuals with higher education attainment have a high propensity to support gender egalitarian views. Ye & Zhao (2018) further confirm that the impact of traditional gender identity on labour participation decreases when the cohort's education attainment increases. Similarly, ed- ucation attainment's effect in enhancing economic well-being has long been established Stiglitz (1975). Provided that egalitarian gender role beliefs and higher incomes could simultaneously be the product of education attainment, it is not surprising to see that higher-income women are less susceptible to traditional gender roles and less willing to diminish their income. Our second explanation is about the financial cost of conforming to gender norms. The smaller magnitude of discontinuity among high-income spouses suggest that less married women are willing to compromise their labour outcomes to gender norms. This could be because they need to discard a larger amount of earnings in achieving that. The high financial cost associated with conforming to gender norms discentivise them in doing so.

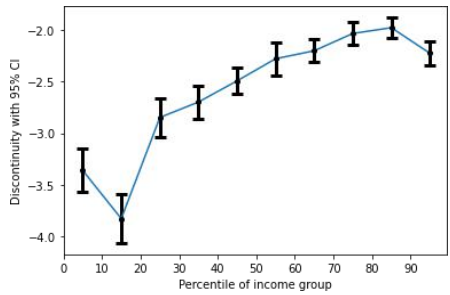

Figure 3: The discontinuity and 95% con- fidence interval for different female income groups.

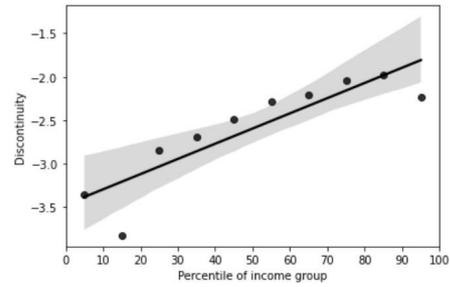

Figure 4: Correlation between the discon- tinuity and income group based on the women's earnings.

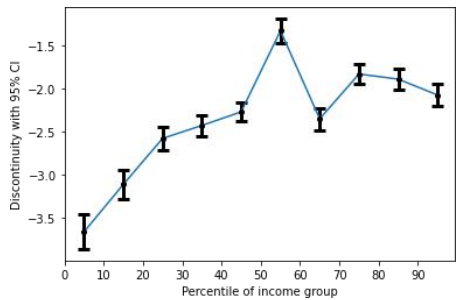

Figure 5: The discontinuity and 95% con- fidence interval for different household in- come groups.

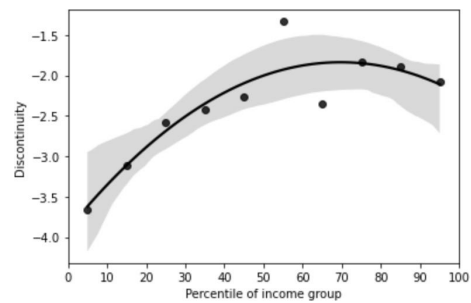

Figure 6: Correlation between the discon- tinuity and income group based on the spousal earnings.

In the following part, we aim to assess an alternative explanation of the discontinuity proposed by Zinovyeva & Tverdostup (2018) in the context of China. They disagree with how Bertrand et al. (2015) use gender identity to explain the discontinuity and instead propose that the discontinuity is caused by the spouses working together. Their key ideas could be summarised as the following: (1) Co-working spouses include couples who are employers working in the same premises or both self-employed. They work in the same industry and are likely to report identical incomes. This drives up the number of equally-earning spouses and creates discontinuity. (2) They find that the discontinuity is insignificant for the newly-weds and only arises gradually along with the marriage. This could be explained as husbands and wives probably start to work together and earn equally only after their cohabitation.

We argue against the co-working hypothesis for the following reasons. First, the questionnaire of CHFS is designed in a way that respondents are only asked to report their yearly incomes after declaring to be employees. Self-employers, business owners and the unemployed are not required to report their incomes, and are hence omitted from our dataset. Thus, the discontinuity could only become larger if self-employers and business owners are included. We also argue that income is not an accurate proxy for spouses' respective economic contributions when couples are business partners or self-employers. This is because despite the same income may be reported, there is still a possibility that one party devotes more time and effort into their economic activities than the other, so he or she can be deemed as the main breadwinner.

Second, both Zinovyeva & Tverdostup (2018) and our datasets do not record whether spouses share the same employers. They first identify spouses both working in the same industry as co-working spouses, provided that Hyatt (2015) concludes spouses in the same industry are very likely to share one employer. Then, they prove their alternative expla- nation by showing no discontinuity among spouses that do not work with each other. To testify their work, we follow the same identification method. Our dataset suggests that more than 40% of couples can be identified as co-working spouses (10,000 out of 21,000). We apply McCrary (2008) test on co-working and non-co-working couples. The results, presented in Table 2 and visualized in Figure 7 to 10, show that the distribution of co-working spouses drops by approximately 300% at the point which wives become the main income contributors. The drop exhibited in the cohort of non-co-working spouses is approximately 100%. The fact that the large drop still exists indicates that spouses work- ing together is unlikely to be the main cause of discontinuity. The results are also robust after excluding all identical-earning couples. In addition, co-working spouses identified by us can potentially include those working in the same industry but different firms. For these spouses, gender norms could still render wives to diminish their income to a level equal to their husbands',leading to the discontinuity. Our current dataset cannot address this problem because information about the employers is not provided. In general, we conclude that co-working cannot fully explain the discontinuity and thus are unable to rule out our gender identity explanation.

Third, Zinovyeva & Tverdostup (2018) argue that the convergence of income usually occurs gradually during the marriage tenure. They suggest that newlyweds are less likely to earn identical income, and their income will likely converge during marriage tenure. The convergence of income could occur in two ways: less-earning wives catch up with husbands and finally earn the same, or vice versa. Our dataset is limited as it does not record at which year the couples start to cohabit or get married. We therefore have to use the average age of couples as a proxy for their length of the marriage. If the co-working hypothesis in Zinovyeva & Tverdostup (2018) is held true, we are supposed to see the discontinuity increases as the age group ascends. We calculate the mean age of each spouse and divide all couples into four quartiles based on that. The results are presented in Table 3. There is no salient or consistent pattern of discontinuity among couples of different age groups. The results are robust after omitting the identical earning couples.

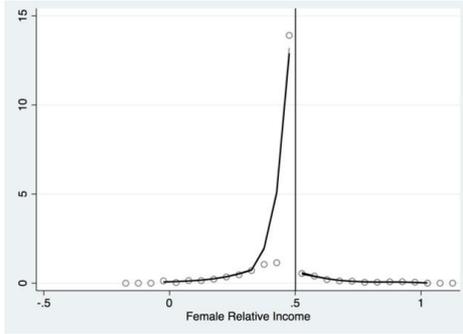

Figure 7: McCrary test for all co-working couples at the right side of the spousal equal earning point (0.501).

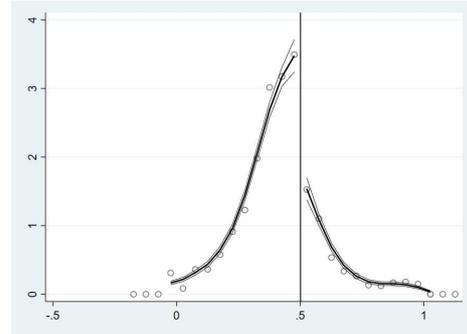

Figure 8: McCrary test for co-working and non-identical earning couples at the right side of the spousal equal earning point (0.501).

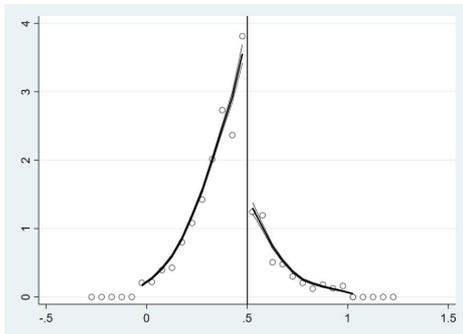

Figure 9: McCrary test for all non-co-working couples at the right side of the spousal equal earning point (0.501).

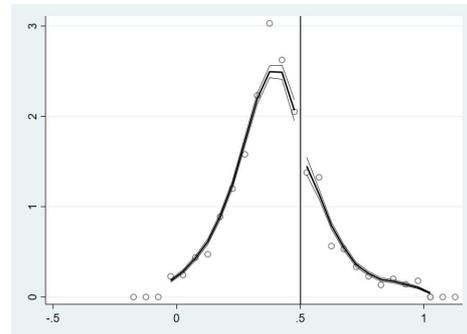

Figure 10: McCrary test for non-co-working and non-identical earning couples at the right side of the spousal equal earn- ing point (0.501).

Table 2: **Magnitude of Discontinuity in Co-working Spouses (Including and Excluding 0.5)**

|  | (1) | | (2) | |
| --- | --- | --- | --- | --- |
| Group | Log Diff. | Obs | Log Diff. | Obs |
| Co-working spouses | -3.480 | 8,344 | -0.945 | 3,219 |
|  | (0.088) |  | (0.087) |  |
| Non co-working spouses | -1.097 | 12,371 | -0.239 | 11,089 |
|  | (0.066) |  | (0.072) |  |

Notes: The first two columns in (1) indicate the results of McCrary test on all females while the third and fourth columns in (2) indicate the results of McCrary test on females that do not earn the same wage as their husbands. Our results suggest that the discontinuity is not caused by spouses working together, and therefore the alternative explanation of Zinovyeva & Tverdostup (2018) is not applicable to China's context.

Table 3: **Magnitude of Discontinuity in Different Age Income Groups**

|  | (1) | | (2) | |
| --- | --- | --- | --- | --- |
| Group | Log Diff. | Obs | Log Diff. | Obs |
| Age below $25^{th}$ pct. | -1.971 | 5,458 | -0.392 | 3,388 |
|  | (0.076) |  | (0.052) |  |
| Age between $25^{th}$ and $50^{th}$ pct. | -1.966 | 5,282 | -0.436 | 3,418 |
|  | (0.074) |  | (0.097) |  |
| Age between $50^{th}$ and $75^{th}$ pct. | -2.136 | 5,164 | -0.480 | 3,377 |
|  | (0.080) |  | (0.104) |  |
| Age above $75^{th}$ pct. | -2.072 | 5,171 | -0.389 | 2,945 |
|  | (0.099) |  | (0.120) |  |

Notes: The first two columns in (1) indicate the results on all females, while in the third and fourth columns in (2) indicates the results of removing couples with identical earnings. Zinovyeva & Tverdostup (2018) suggest that there is no discontinuity among newly cohabited couples, and discontinuity only arises during their tenure of marriage during which they start to work together. Because we do not have specific data on the number of years of marriage, we use the age of females to proxy for the tenure of marriage. We divide our sample into four groups based on their ages. As our result suggest, there is no substantial variation on the size of discontinuity among different cohorts based on ages.

# 5 Regression

## 5.1 Baseline model — fixed effect regression with time lag

We estimate the impact of gender norms on females' labour supply by using fixed effect regression with time lag. We choose to use a similar identification strategy as Sprengholz et al. (2020) to illustrate that married women in China would diminish their labour market outcome to avoid a situation where she would earn more than her husband. If a female earns more than her husband initially but her labour outcome is diminished in the following years, we could argue that gender norms affects females' labour outcomes. Our baseline regression is specified as:

$$WifeEmployed_{i,t} = \beta_1 \cdot WifeEarnsMore_{i,t-1} + \beta_2 \cdot \mathbf{X}_{i,t} + \delta_n + \mu_t^R + s_{i,t} \tag{4}$$

$$LogIncome_{i,t} - LogIncome_{i,t-1} = \beta_1 \cdot WifeEarnsMore_{i,t-1} + \beta_2 \cdot \mathbf{X}_{i,t} + \delta_n + \mu_t^R + s_{i,t} \tag{5}$$

$$WeekWorkHr_{i,t} - WeekWorkHr_{i,t-1} = \beta_1 \cdot WifeEarnsMore_{i,t-1} + \beta_2 \cdot \mathbf{X}_{i,t} + \delta_n + \mu_t^R + s_{i,t} \tag{6}$$

Three specifications are given in this chapter. The variable of interest is the same: WifeEarnsMore$_{i,t-1}$, a binary variable indicating whether the female i earns more than her husband in the previous wave. Through the three specifications, we investigate whether females whose earnings exceed their husbands would reduce their labour outcomes to conform to gender norms by leaving their employment posts, reducing their incomes, or working for fewer hours. If our parameter of interest is negative and signif- icant, we can conclude that females who earn more than their husbands are more likely to distort their labour outcomes than others earning less than their spouses.

$X_{i,t}$ represents the additional demographic controls for individual i at time t. The controls include, first, ages of the respondents, which are represented by the variable Age, and its square Age square. Second, we generate a binary variable to indicate whether female i has received an education equal or higher than high school or technical college, with 1 representing the woman is highly educated and 0 otherwise. Third, we generate a binary variable indicating whether female i is affiliated to the Chinese Communist Party (CCP), with 1 being a CCP member and 0 otherwise. These variables are controlled because they could also impact female labour outcomes. For example, being a CCP member means one is more likely to have access to high-ranked officials and be promoted, therefore making this female reluctant to diminish her labour outcomes.

In addition, we construct two fixed effect groups to conduct fixed effect regression. The first group includes year and province, and the second group includes the industry in which the married woman i is working. Our concern is that wife out-earning husbands might be conditional on unobserved characteristics of different regions and industries, making our variable of interest endogenous. Therefore, we apply two fixed effect groups to partial out these effects.

The results of the baseline model are presented in Table 4. Our results provide concrete evidence on how a wife earning more than her husband at t−1 would impact her future labour participation decision as the coefficients of $WifeEarnsMore_{i,t-1}$ are all negative and significant. This could prove the central thesis of this research that when a married woman earns more than her husband, she would probably adjust her labour outcomes to conform to gender norms. Specifically, when the dependent variable is change in logged income, the coefficient is approximately - 0.2 under 1% significance level. It suggests that she could approximately diminish 20 per cent of her income on average as an attempt to conform to gender norms. For the change in weekly working hours, the coefficient is approximately -1 under 1% and 5% significance levels, without and with fixed effects treatment, respectively. This indicates that she could reduce her weekly working time by 1 hour.

Table 4: **Result of the Baseline Model**

| | Employe | | Change in WorkHr | | Change in LogInc | |
|---|---|---|---|---|---|---|
| | (1) | (2) | (3) | (4) | (5) | (6) |
| $WifeEarnsMore_{t-1}$ | 0.0113** | 0.00926* | -0.182*** | -0.178*** | -1.054*** | -0.937** |
| | (0.00550) | (0.00502) | (0.0298) | (0.0291) | (0.389) | (0.391) |
| Age | 0.00530*** | 0.00514*** | 0.0221** | 0.0146 | 0.137 | 0.123 |
| | (0.00187) | (0.00172) | (0.0102) | (0.00996) | (0.138) | (0.140) |
| Age² | -7.95e-05*** | -6.54e-05*** | -0.000187 | -0.000149 | -0.00145 | -0.00132 |
| | (2.17e-05) | (1.99e-05) | (0.000117) | (0.000115) | (0.00161) | (0.00163) |
| High | 6.99e-06 | 0.0208*** | 0.0477* | 0.0288 | -0.0284 | 0.0761 |
| | (0.00524) | (0.00497) | (0.0283) | (0.0287) | (0.376) | (0.395) |
| Hukoutype agri | -0.00380 | -0.00825* | 0.00263 | -0.0302 | 0.184 | 0.0944 |
| | (0.00519) | (0.00493) | (0.0281) | (0.0285) | (0.374) | (0.393) |
| Ccpmember yes | 0.0101 | 0.00762 | -0.0802** | 0.0108 | 0.420 | 0.568 |
| | (0.00615) | (0.00572) | (0.0330) | (0.0328) | (0.428) | (0.440) |
| Constant | 0.894*** | 0.864*** | -0.560** | -0.305 | -3.118 | -2.852 |
| | (0.0407) | (0.0374) | (0.221) | (0.217) | (2.980) | (3.014) |
| Observations | 6,235 | 6,232 | 6,121 | 6,118 | 5,728 | 5,725 |
| R-squared | 0.010 | 0.213 | 0.010 | 0.102 | 0.002 | 0.040 |
| Province Year FE | No | Yes | No | Yes | No | Yes |
| Industry FE | No | Yes | No | Yes | No | Yes |

Notes: This table reports how "a wive earns more than her husband at $t-1$" causes changes in her labour outcome at $t$. The dependent variable in Column (1) and (2) "*Employed*", which represents the employment status of wives at time $t$. The dependent variable in Column (3) and (4) is "*Change in LogInc*", which represents the wives' log income between $t$ and $t-1$. The dependent variable in Column (5) and (6) is "*Change in WorkHr*", which represents the changes of weekly working hours of between $t$ and $t-1$. Column (1),(3),(5) report the coefficients from OLS regressions, while Column (2),(4),(6) report the coefficients from fixed effect regressions. The effect time includes year and province, while the second one includes the industry in which a married woman $i$ is working. The variable of interest, "$WifeEarnsMore_{t-1}$" is a binary variable that indicates if wives earn more than husbands at $t-1$. Standard errors in parentheses: *** p<0.01, ** p<0.05, * p<0.1

## 5.2 Heterogeneity studies

We note that the magnitude of discontinuity varies for couples with different education levels and income groups. Our conclusions in Chapter 4 are that wives with less education attainment are more likely to distort their labour outcomes and avoid the circumstances that they earn more than their husbands. It is also concluded in Chapter 4 that couples from low-income groups are more likely to distort their labour outcomes and reduce their earnings. We want to testify to this heterogeneity in this section.

### 5.2.1 Education level

The prior section has told us that the magnitude of discontinuity at the 0.501 point decreases when the female is highly educated. This characteristic is also observed by Bertrand et al. (2015) using a dataset from the United States. We divide our dataset into two groups, depending on whether their educational attainments are higher than or equivalent to high schools (including technical schools). Since the baseline models have indicated that women will not distort their employment status when they earn more than husbands, we only examine how distortion of income and weekly working hours vary among groups with different educational attainments. We use the following specifications:

$$\begin{aligned}
LogIncome_{i,t} - LogIncome_{i,t-1} =& \beta_1 \cdot WifeEarnsMore_{i,t-1} + \\
& \beta_2 \cdot High\ edu + \beta_3 \cdot Low\ edu + \\
& \beta_4 \cdot WifeEarnsMore_{i,t-1} \times High\ edu + \\
& \beta_5 \cdot WifeEarnsMore_{i,t-1} \times Low\ edu + \\
& \beta_6 \cdot \mathbf{X}_{i,t} + \delta_n + \mu_t^R + s_{i,t}
\end{aligned} \quad (7)$$

$$\begin{aligned}
WeekWorkHr_{i,t} - WeekWorkHr_{i,t-1} =& \beta_1 \cdot WifeEarnsMore_{i,t-1} + \\
& \beta_2 \cdot High\ edu + \beta_3 \cdot Low\ edu + \\
& \beta_4 \cdot WifeEarnsMore_{i,t-1} \times High\ edu + \\
& \beta_5 \cdot WifeEarnsMore_{i,t-1} \times Low\ edu + \\
& \beta_6 \cdot \mathbf{X}_{i,t} + \delta_n + \mu_t^R + s_{i,t}
\end{aligned} \quad (8)$$

We compare the coefficients before the two interaction terms to identify the difference of distortion of labour outcomes between couples with different educational attainments. Note that the absolute value of coefficient before interaction term of less educational attainment is larger than the coefficient before the interaction terms of more educational attainment when it comes to distortion of income. Interestingly, we notice that higher educated women decrease their weekly working hours by a slightly larger magnitude.

Table 5: **Change in Wife's Labour Outcomes from Different Education Groups When She Earns More Than Her Husband.**

| Variable | Change in LogInc | Change in WorkHr |
|---|---|---|
| $WifeEarnsMore_{t-1} \times High\_edu$ | -0.150*** | -0.982** |
| | (0.0353) | (0.472) |
| $WifeEarnsMore_{t-1} \times Low\_edu$ | -0.239*** | -0.839 |
| | (0.0513) | (0.699) |
| Prob>F | 0.1508 | 0.8653 |
| Observation | 6,118 | 5,725 |
| Province Year FE | Yes | Yes |
| Industry FE | Yes | Yes |

Notes: We study the heterogeneous impact of gender views on labour outcomes between cohorts with different educational attainment. The results have shown that the married women with less educational attainment could more likely to reduce their labour outcomes than those with educational level above high school. Standard errors in parentheses: *** p<0.01, ** p<0.05, * p<0.1

However, no concrete conclusions can be drawn at this stage because the coefficient before the interaction term of lower educational attainment is insignificant.

### 5.2.2 Income group

We are interested to know whether a female's social-economic status could affect her decision on whether she will distort her labour outcomes when over-earning her husband. Hence, we divide the whole sample into four groups based on women's individual income and the household income respectively. Four interactions are subsequently constructed as the interaction between the $IncomeGroup\text{"}n\text{"}$ dummy and $WifeEarnsMore_{t-1}$ dummy. We therefore construct the following specifications:

$$\begin{aligned}
LogIncome_{i,t} - LogIncome_{i,t-1} =& \beta_1 \cdot WifeEarnsMore_{i,t-1} + \\
& \beta_2 \cdot IncomeGroup1 + \beta_3 \cdot IncomeGroup2 + \\
& \beta_4 \cdot IncomeGroup3 + \beta_5 \cdot IncomeGroup4 + \\
& \beta_6 \cdot WifeEarnsMore_{i,t-1} \times IncomeGroup1 + \\
& \beta_7 \cdot WifeEarnsMore_{i,t-1} \times IncomeGroup2 + \\
& \beta_8 \cdot WifeEarnsMore_{i,t-1} \times IncomeGroup3 + \\
& \beta_9 \cdot WifeEarnsMore_{i,t-1} \times IncomeGroup4 + \\
& \beta_{10} \cdot \mathbf{X}_{i,t} + \delta_n + \mu_t^R + s_{i,t}
\end{aligned} \quad (9)$$

$$\begin{aligned}
WeekWorkHr_{i,t} - WeekWorkHr_{i,t-1} = & \beta_1 \cdot WifeEarnsMore_{i,t-1} + \\
& \beta_2 \cdot IncomeGroup1 + \beta_3 \cdot IncomeGroup2 + \\
& \beta_4 \cdot IncomeGroup3 + \beta_5 \cdot IncomeGroup4 + \\
& \beta_6 \cdot WifeEarnsMore_{i,t-1} \times IncomeGroup1 + \\
& \beta_7 \cdot WifeEarnsMore_{i,t-1} \times IncomeGroup2 + \\
& \beta_8 \cdot WifeEarnsMore_{i,t-1} \times IncomeGroup3 + \\
& \beta_9 \cdot WifeEarnsMore_{i,t-1} \times IncomeGroup4 + \\
& \beta_{10} \cdot \mathbf{X}_{i,t} + \delta_n + \mu_t^R + s_{i,t}
\end{aligned} \quad (10)$$

For both dependent variables, the coefficients decrease in absolute value when the income group ascends. This has indicated that high-earning women or women from high- income households are more susceptible to gender norms and are less likely to reduce their income or weekly working hours when they earn more than their husbands.

**Table 6: Change in Wife's Labour Outcomes from Different Income Groups When She Earns More Than Her Husband.**

| Variable | (1) Change in LogInc | Change in WorkHr | (2) Change in LogInc | Change in WorkHr |
|---|---|---|---|---|
| $IncomeGroup1 \times WifeEarnsMore_{t-1}$ | -1.042*** | -2.229** | -0.685*** | -1.203 |
|  | (0.0707) | (0.949) | (0.0606) | (0.822) |
| $IncomeGroup2 \times WifeEarnsMore_{t-1}$ | -0.112** | -0.770 | -0.100* | -0.0953 |
|  | (0.0550) | (0.768) | (0.0532) | (0.724) |
| $WifeEarnsMore_{t-1} \times IncomeGroup3$ | -0.108** | -0.902 | -0.0972* | -1.096 |
|  | (0.0518) | (0.708) | (0.0530) | (0.711) |
| $WifeEarnsMore_{t-1} \times IncomeGroup4$ | 0.101** | -0.692 | 0.0845* | -1.124* |
|  | (0.0438) | (0.593) | (0.0499) | (0.677) |
| $WifeEarnsMore_{t-1}$ | 0.000 | 0.553 | 0.000 | 0.662 |
|  | 6,118 | 5,725 | 6,118 | 5,725 |
| Prob>F | Yes | Yes | Yes | Yes |
|  | s | s | s | s |

Notes: In the first two columns under (1), Income Group 1 refers to females whose individual income are below 25 percentile, Group 2 refers to females whose individual income are between 25 and 50 percentile, Group 3 refers to females whose individual income are between 50 and 75 percentile, and Group 4 refers to females whose individual income are above 75 percentile. In the next two columns under (2), females are grouped using the household income (the sum of variables of interest are the coefficients before the interaction terms between the income group dummy and "$WifeEarnsMore_{t-1}$." Two fixed effects are applied columns. Standard errors in parentheses: *** p<0.01, ** p<0.05, *

# 6 Robustness Check

To check the robustness of our proposition, we first change the definition of a wife out-earning husband. Two reasons arguably demonstrate why changes of definition are needed. First, husbands could feel their breadwinner status being challenged both when their wives actually earn more than them and when the wives' income are approaching but have not exceeded theirs. As a result, they might start to negotiate with their wives and influence their wives to diminish labour outcomes. Second, for spouses earning similar wages, wives out-earning husbands could be a random event. In these cases, wife out-earning husbands could be an intrinsically non-persistent phenomenon, yet when wives earn less in the future, these could still be interpreted as female diminishing labour outcomes to echo gender norms. Therefore, we have refined the variable of interest wife out-earn in the following ways: (1) wife's relative income exceeds 0.45 (2) wife's relative income exceeds 0.55.

The results are presented in Table A6 and A7. The results are consistent with our baseline model. Therefore, we argue that our baseline model is still robust.

Murray-Close & Heggeness (2018) have shown that among couples where wives earn more than husbands, wives are more likely to under-report their income to a level equal to the husband. Sprengholz et al. (2020) also argue that couples could report that they are equal earning although their actual earnings could differ. We therefore delete all observations of equal earning husbands and wives and the results are presented in Table A8. Similarly, we further restrict our sample to those couples aged between 25 and 59. This group is more likely to include couples who have entered the labour force yet are not retired. The test results in Table A9 show that our central thesis is still robust.

# 7 Conclusion

Literature and survey results have shown that a traditional gender view that males should be the major breadwinner of the household. We have empirically studied how females conform to traditional gender views by distorting their labour outcomes. In the first part of our research, we employ the discontinuity test to the right of 0.5 of a female's relative income. The large discontinuity indicates that females have an incentive to avoid their income exceeding their husbands'. The magnitude obtained in this research is larger than the results obtained in Western nations such as the United States and Germany. We also assess the validity of Zinovyeva & Tverdostup (2018)' alternative explanation that the discontinuity was caused by husbands and wives working together. We conclude that their explanation is not applicable in China's context. In the second part of our research, we employed the fixed effect regression with time lag to quantify the distortion of future labour outcomes when females earn more than their husbands. We conclude that a wife earning more than her husband will not affect her labour participation decision. However, there is sufficient evidence to demonstrate that women who earn more than their husbands will decrease their income and working hours. We also explore the heterogeneous impact of gender views on the distortion of gender views in cohorts differentiated by educational attainment and income. We conclude that females with less educational attainment and less income are more susceptible to traditional gender views.

There are some limitations in our research. Firstly, our dataset, China Household Financial Survey, is collected biannually. This means that we can only use the relative earning status to predict the change of labour outcomes in the following two years. The longer time interval could introduce more omitted variables that are not included in our specifications. Secondly, Murray-Close & Heggeness (2018) suggests that women who just earn more than husbands are more likely to under-report their earnings, resulting in the spike of frequency where husbands and wives earn the same wage. The intrinsic bias of the data could affect the validity of our conclusion.

Lastly, our research possesses policy implications. This could serve as a reminder to policymakers that fully-employed married women could decrease their labour outcomes to conform to traditional gender views. Currently, policymakers in China are mainly con- cerned about the decreasing female labour participation rate (FLFP) because its declining supply of labour could negatively affect productivity and capital utilisation. Our research provides a new perspective to assess how traditional gender views affect labour produc- tivity: other than those women who leave employment posts and result in the declining trend of FLFP rate, those who stay in employment posts could still compromise their labour outcomes and cause a less efficient use of labour resources. To policymakers who want to optimise the labour structure by optimising labour outcomes of females, they could consider policies that reshape gender views of the public so that people could be more friendly and acceptable to women who earn more than their husbands.

# A  Appendix

Table A1: **Data Summary for All Couples Participating in CHFS**

| Variable | Obs | Mean | SD |
|---|---|---|---|
| *Wife's earning (compared to her husband)* | | | |
| *WifeEarnsMore$_{i,t}$* (earn more=1, o/w=0) | 21,069 | 0.159 | 0.366 |
| *WifeEarnsEqual$_{i,t}$* (earn equal=1, o/w=0) | 21,069 | 0.371 | 0.483 |
| *WifeEarnsLess$_{i,t}$* (earn less=1, o/w=0) | 21,069 | 0.469 | 0.499 |
| *Income* | | | |
| Wife's yearly income after tax | 21,046 | 31,758 | 40,597 |
| *Wife_Relative_Inc$_{i,t}$* | 20,897 | 0.445 | 0.145 |
| *Working time* | | | |
| Wife's weekly working time | 19,360 | 42.194 | 10.906 |

Table A2: **Data Summary for Couples Participating in CHFS for More Than One Wave**

| Variable | Obs | Mean | SD |
|---|---|---|---|
| *Employed* | 14,217 | 0.959 | 0.199 |
| Yearly income after tax | 14,199 | 30,397 | 41,135 |
| Weekly working hours | 13,097 | 42.135 | 42.136 |
| *WifeEarnsMore$_{i,t}$* | 14,217 | 0.166 | 0.372 |
| *Age* | 14,216 | 42.197 | 9.0186 |
| *Age$^2$* | 14,216 | 1861.9 | 790.65 |
| *High_edu* | 14,217 | 0.549 | 0.498 |
| *CCP_member* | 14,217 | 0.124 | 0.329 |
| *Agri_residency* | 14,217 | 0.382 | 0.486 |

Notes: "*High_edu*" is a binary variable with 1 representing female "*i*" obtained a diploma equal to or higher than high school, and 0 otherwise. "*CCP member*" is a binary variable with 1 representing female "*i*" joins the Communist Party of China, and 0 otherwise. "*Agri residency*" is a binary variable with 1 representing female "*i*" holds agricultural "hukou" (which indicates she is living in rural regions), and 0 otherwise.

Table A3: **McCrary Test Results on Different Income Groups of Married Women**

| Group | Log Diff. | SE | Obs |
|---|---|---|---|
| Below $10^{th}$ pct. | -3.359 | 0.2102 | 2,103 |
| Between $10^{th}$ and $20^{th}$ pct. | -3.825 | 0.2359 | 2,477 |
| Between $20^{th}$ and $30^{th}$ pct. | -2.848 | 0.1845 | 1,734 |
| Between $30^{th}$ and $40^{th}$ pct. | -2.698 | 0.1607 | 1,874 |
| Between $40^{th}$ and $50^{th}$ pct. | -2.494 | 0.1292 | 2,608 |
| Between $50^{th}$ and $60^{th}$ pct. | -2.280 | 0.1628 | 1,710 |
| Between $60^{th}$ and $70^{th}$ pct. | -2.201 | 0.1092 | 2,360 |
| Between $70^{th}$ and $80^{th}$ pct. | -2.036 | 0.1064 | 2,384 |
| Between $80^{th}$ and $90^{th}$ pct. | -1.980 | 0.0966 | 2,428 |
| Above $90^{th}$ pct. | -2.227 | 0.1138 | 1,791 |

Notes: We divide the sample into ten income groups based on females' incomes and carry out McCrary test on the right side of spousal equal earning point. We cannot exactly divide the sample into ten equal-sized groups because defaulting lines of each tenfold usually consist of many women with the same income, and we have to include those women into that income group. For example, the married women who earns more than 6,200 by less than or equal to 12,000 are included into 10-20 percentile. The fact that we include all the women in the upper limit cause the difference in numbers of observations in each income groups. Our results suggest that the magnitude of discontinuity decreases among high income cohorts. This indicates that women earning higher income are less susceptible to gender views and less likely to decrease their labour outcomes.

Table A4: **McCrary Test Results on Different Income Groups based on Household Income**

| Group | Log Diff. | SE | Obs |
|---|---|---|---|
| Below $10^{th}$ pct. | -3.660 | 0.1990 | 2,244 |
| Between $10^{th}$ and $20^{th}$ pct. | -3.112 | 0.1701 | 2,020 |
| Between $20^{th}$ and $30^{th}$ pct. | -2.579 | 0.1363 | 2,062 |
| Between $30^{th}$ and $40^{th}$ pct. | -2.428 | 0.1184 | 2,276 |
| Between $40^{th}$ and $50^{th}$ pct. | -2.271 | 0.1047 | 2,852 |
| Between $50^{th}$ and $60^{th}$ pct. | -1.332 | 0.1375 | 1,582 |
| Between $60^{th}$ and $70^{th}$ pct. | -2.351 | 0.1280 | 2,126 |
| Between $70^{th}$ and $80^{th}$ pct. | -1.832 | 0.1103 | 2,142 |
| Between $80^{th}$ and $90^{th}$ pct. | -1.892 | 0.1187 | 1,848 |
| Above $90^{th}$ pct. | -2.076 | 0.1287 | 2,126 |

Notes: We divide the sample into ten income groups based on the spousal total income and carry out McCrary test on the right side of the spousal equal earning point. The reason of differences in the number of observations in each income groups is explained in the text and note in Table 4. Our results suggest that the graph of the magnitude of discontinuity against the spousal income shows an inverted U-shape.

Table A5: **Density Discontinuity Estimates for Working Couples (Restricted to Couples with at Least Two Waves of Survey Result)**

| Group | Log Diff. | SE | Obs |
|---|---|---|---|
| All dual-earning couples | -2.299 | 0.0487 | 14,217 |
| Non-identical earning couples | -3.112 | 0.1701 | 9,097 |

Notes: Different from Table 1, we omit all couples who have only participate in one wave of CHFS survey from our dataset. The results are shown above.

Table A6: **Robustness check 1:** (*WifeEarnsMore* = 1 **if female income percentage exceeds 45%**)

| Variable | Employed | Change in LogInc | Change in WorkHr |
|---|---|---|---|
| *WifeEarnsMore*$_{t-1}$ | 0.00821* | -0.244*** | -0.388 |
|  | (0.00435) | (0.0245) | (0.424) |
| Observation (Household-year) | 10442 | 10442 | 9428 |
| Observation | 5,221 | 5,221 | 4,714 |
| Province Year FE | Yes | Yes | Yes |
| Industry FE | Yes | Yes | Yes |

Notes: Standard errors in parentheses: *** p<0.01, ** p<0.05, * p<0.1

Table A7: **Robustness check 2:** (WifeEarnsMore = 1 **if female income percentage exceeds 55%**)

| Variable | Employed | Change in LogInc | Change in WorkHr |
|---|---|---|---|
| *WifeEarnsMore*$_{t-1}$ | -0.00426 | -0.243*** | -1.596** |
|  | (0.00708) | (0.0401) | (0.707) |
| Observation (Household-year) | 10442 | 10442 | 9428 |
| Observation | 5,221 | 5,221 | 4,714 |
| Province Year FE | Yes | Yes | Yes |
| Industry FE | Yes | Yes | Yes |

Notes: Standard errors in parentheses: *** p<0.01, ** p<0.05, * p<0.1

Table A8: **Robustness check 3: (Drop all equal earning spouses)**

| Variable | Employed | Change in LogInc | Change in WorkHr |
|---|---|---|---|
| *WifeEarnsMore*$_{t-1}$ | 0.00252 | -0.0838 | -1.695* |
|  | (0.00699) | (0.0536) | (0.897) |
| Observation (Household-year) | 7972 | 4360 | 4114 |
| Observation | 3,986 | 2,180 | 2,057 |
| Province Year FE | Yes | Yes | Yes |
| Industry FE | Yes | Yes | Yes |

Notes: Standard errors in parentheses: *** p<0.01, ** p<0.05, * p<0.1

Table A9: **Robustness check 4: (Restrict to those aged between 25 and 56)**

| Variable | Employed | Change in LogInc | Change in WorkHr |
|---|---|---|---|
| *WifeEarnsMore*$_{t-1}$ | 0.00638 | -0.105** | -1.595** |
|  | (0.00629) | (0.0491) | (0.780) |
| Observation (Household-year) | 10482 | 5688 | 5410 |
| Observation | 5,241 | 2,844 | 2,705 |
| Province Year FE | Yes | Yes | Yes |
| Industry FE | Yes | Yes | Yes |

Notes: Standard errors in parentheses: *** p<0.01, ** p<0.05, * p<0.1